\documentclass[twocolumn,preprintnumbers,amssymb]{revtex4}
\begin{document} 
\newcommand{\sech}{{\rm sech}} 
\newcommand{\csch}{{\rm csch}} 
\newcommand{\beq}{\begin{eqnarray}}
\newcommand{\eeq}{\end{eqnarray}} 
\newcommand{\nn}{\nonumber} 
\def\ltap{\ \raise.3ex\hbox{$<$\kern-.75em\lower1ex\hbox{$\sim$}}\ } 
\def\gtap{\ \raise.3ex\hbox{$>$\kern-.75em\lower1ex\hbox{$\sim$}}\ } 
\def\CO{{\cal O}} 
\def\CL{{\cal L}} 
\def\CM{{\cal M}} 
\def\tr{{\rm\ Tr}}\ 
\def\CO{{\cal O}} 
\def\CL{{\cal L}} 
\def\CM{{\cal M}} 
\def\tr{{\rm\ Tr}} 
\newcommand{\bel}[1]{\be\label{#1}} 
\def\al{\alpha} 
\def\bt{\beta} 
\def\eps{\epsilon} 
\def\mn{{\mu\nu}} 
\newcommand{\rep}[1]{{\bf #1}} 
\def\be{\begin{equation}} 
\def\ee{\end{equation}} 
\def\bea{\begin{eqnarray}} 
\def\eea{\end{eqnarray}} 
\newcommand{\eref}[1]{(\ref{#1})} 
\newcommand{\Eref}[1]{Eq.~(\ref{#1})} 
\newcommand{\gsim}{ \mathop{}_{\textstyle \sim}^{\textstyle >} } 
\newcommand{\lsim}{ \mathop{}_{\textstyle \sim}^{\textstyle <} } 
\newcommand{\vev}[1]{ \left\langle {#1} \right\rangle } 
\newcommand{\bra}[1]{ \langle {#1} | } 
\newcommand{\ket}[1]{ | {#1} \rangle } 
\newcommand{\ev}{ {\rm eV} } 
\newcommand{\kev}{ {\rm keV} } 
\newcommand{\Mev}{ {\rm MeV} } 
\newcommand{\gev}{ {\rm GeV} } 
\newcommand{\tev}{ {\rm TeV} } 
\newcommand{\mev}{ {\rm meV} } 
\newcommand{\ma}{m^2_{\rm atm}} 
\newcommand{\ml}{m^2_{\rm LSND}} 
\newcommand{\tl}{\theta_{\rm LSND}} 
\newcommand{\ms}{m_\odot^2} 
\newcommand{\cta}{c_{\rm a}} 
\newcommand{\cts}{c_\odot} 
\newcommand{\sta}{s_{\rm a}} 
\newcommand{\sts}{s_\odot} 
\newcommand{\ctm}{c_{\rm m}} 
\newcommand{\stm}{s_{\rm m}} 
\newcommand{\mpl}{M_{Pl}}

\title{Extended Anomaly Mediation and New Physics at 10 TeV}

\author{Ann E. Nelson} 
\author{Neal J. Weiner} 
\affiliation{Department of Physics, Box 1560, University of Washington, Seattle, WA 98195-1560, USA} 
\date{\today} 





\begin{abstract} 
In the MSSM, an unfortunate prediction of minimal anomaly mediated 
supersymmetry breaking is that the slepton masses squared are negative. 
This problem is particularly intractable because of the insensitivity of 
anomaly mediation to ultraviolet physics. In this paper we note that tree 
level couplings to the conformal compensator in the K\"ahler potential give 
10 TeV as a natural mass scale for physics beyond the MSSM, and, moreover, 
that the SUSY breaking effects from physics at this scale do not 
generically decouple from the low-energy spectrum. We consider particular 
extensions, including the effects of vector-like matter at 10 TeV, 
and a specific model in which the leptons are placed in a triplet of 
an asymptotically free $SU(3)$. We find that the features of minimal 
anomaly mediation are not a robust prediction of the general framework, 
and that the problem of negative slepton masses squared can easily be avoided.

\end{abstract}  
\maketitle

\section{Introduction:  The Supersymmetric Flavor Problem, and Anomaly Mediation}
\label{sec:intro} 
Extending the standard model of particle physics into a supersymmetric theory is attractive for many reasons, most notably stabilization of the electroweak scale against radiative corrections, and coupling constant unification. A significant drawback to supersymmetric theories is our lack of understanding of supersymmetry breaking. A  phenomenological parametrization of the soft supersymmetry breaking effects in the minimal supersymmetric extension leads to 104 new free parameters. Most of this parameter space is ruled out by low energy constraints on flavor changing neutral currents, lepton number violation, and  CP violation. 
The supersymmetric flavor problem is how to explain the  absence of indirect evidence for supersymmetry via flavor changing neutral currents and lepton flavor violation. The supersymmetric CP problem is why virtual superpartner exchange has not led to  CP violation in conflict with experiment. 

In the simplest  solutions to the supersymmetric flavor and CP  problems the superpartner masses are  insensitive to  Planck scale physics, and can be predicted.   Such UV insensitivity can arise  if the supersymmetry breaking scale is low, as in low energy gauge mediation, where the fundamental supersymmetry breaking scale is between 10 TeV and $10^9$ GeV and the gravitino mass $m_{3/2}$ is   between $10^{-3}$ eV and 1 GeV \cite{Dine:1993yw,Dine:1995vc,Dine:1996ag}.  Typically  Planck scale physics gives contributions to soft supersymmetry breaking masses of order $m_{3/2}$, and so with a low gravitino mass the effects from the Planck scale are negligible. This line of reasoning implies  that the gravitino should be the lightest superpartner (LSP), and  the LSP would not be a candidate for cold dark matter.

Recently there have been several interesting proposals for UV insensitivity of superpartner masses   with $m_{3/2}$ as high as 10 TeV \cite{Randall:1998uk,Fox:2002bu}, or even higher \cite{Luty:2002ff}. Randall and Sundrum \cite{Randall:1998uk} proposed that the hidden supersymmetry breaking sector and the visible supersymmetric extension of the standard model should live in separate (3+1) dimensional subspaces  (known as 3-branes) embedded in extra dimensions. They also proposed that if the bulk contains only supergravity, and the 3-branes are far apart in units of the Planck length,  then the visible sector learns about supersymmetry breaking only via  the breaking of conformal invariance\cite{Randall:1998uk,Giudice:1998xp}. In this ``anomaly mediated supersymmetry breaking'' (AMSB) scenario, the susy breaking parameters at any given scale can be found from an exact solution to the renormalization group equations,  known as the anomaly mediated trajectory. 
Heavy particle threshold corrections will  maintain the soft masses  and couplings on this trajectory, provided the masses of the heavy particles come from supersymmetric terms. Since the superpartner masses are a loop factor below  $m_{3/2}$, and since the weak scale can be determined from the effective supersymmetry breaking scale in the MSSM, a gravitino mass of  order 10 TeV will give the weak scale to be of order 100 GeV.
Thus to leading order in supersymmetry breaking, all the soft supersymmetry breaking parameters at any scale can be computed in terms of the gravitino mass, tree level violation of conformal invariance,  and the beta functions and anomalous dimensions, and is  insensitive to UV physics.

 This predictive scenario would be extremely attractive were it not for the fact that in the Minimal Supersymmetric Standard Model (MSSM), the slepton masses squared are predicted to be negative. Another potential problem with AMSB is that it is not known how to realize this scenario from string theory, and generic  extra  
dimensional models do not realize its predictions---simply separating the 3-branes is insufficient \cite{Anisimov:2001zz}.   However, recently, Luty and Sundrum have shown that AMSB may  be obtained from four dimensional theories with a supersymmetry breaking sector which is embedded in a superconformal sector with certain properties \cite{Luty:2001jh,Luty:2001zv}, so a special extra dimensional set up is  unnecessary for AMSB.

 One  appealing solution to the negative slepton mass squared problem is ``deflected anomaly mediation''  (DAM) \cite{Pomarol:1999ie}, in which   some particles with nontrivial standard model charges obtain large masses from additional light  ($< m_{3/2}$) singlets with large vevs. Integrating  out heavy fields  leads to a  ``deflection'' from the  trajectory at a high energy scale. 
DAM  typically retains some UV sensitivity as it requires running  the soft parameters from the threshold to the weak scale. Flavor violation below the threshold would generically reintroduce the flavor problem. 
There are a number of other viable solutions  to the flavor problem with a gravitino which is heavier than 100 GeV.  Most \cite{Kaplan:1999ac, Chacko:1999am,Chacko:1999mi,  Kaplan:2000jz,Chacko:2000wq,Nomura:2001ub,Chacko:2001jt,Nelson:2001ji}  are  sensitive to  physics  up to some high scale, but not to the Planck scale. 

Solutions to the negative slepton mass squared problem which retain UV insensitivity may be found in refs.  \cite{Jack:2000cd,Jack:2000nt,Arkani-Hamed:2000xj,Carena:2000ad}. These models depart from the anomaly mediated framework by additional   supersymmetry breaking effects  which lie on a different, but UV insensitive, trajectory. Other UV insensitive solutions include \cite{Allanach:2000gu,Katz:1999uw}.

In this paper we will explore anomaly mediation in  theories which go beyond the MSSM at the 10 TeV mass scale, where nondecoupling effects are important. We will argue that   10 TeV is a natural scale for    physics beyond the MSSM.  Since anomaly mediation is still the dominant  contribution to supersymmetry breaking parameters, we dub this framework Extended Anomaly Mediation (EAM). In the EAM framework, for any given  extension of the MSSM,  all supersymmetry breaking parameters can be computed in terms of the gravitino mass and supersymmetric parameters.

\section{Extended Anomaly Mediation}
We start with the following assumptions about the effective theory of  physics below the Planck scale:
\begin{itemize}
\item Anomaly mediation is the dominant source of supersymmetry breaking in the visible sector. 
\item Above $10\ \tev$ all soft supersymmetry breaking terms are on the anomaly mediated trajectory. 
\item The visible sector is some general extension of the MSSM. Typically this will include some new  fields charged under the SM gauge group whose masses are not prevented by a gauge symmetry, but which are nevertheless  light. 
\item No additional mass scales in the vicinity of the weak scale other than $m_{3/2}$ should be input by hand, as it would seem to be overly coincidental for a theory to produce several  mass scales of disparate origin so close together.
\item For  at least some of the new fields, tree level K\"ahler potential terms are allowed  which will generate  masses of $O(m_{3/2})$ from the coupling to $\phi$, the conformal compensator \cite{Giudice:1988yz, Randall:1998uk,Nomura:2001ub}.  
\end{itemize}

We shall see that the resulting low energy superpartner spectrum is not that of minimal anomaly mediation, and we can  easily generate spectra with positive slepton mass squared. 
We  begin by reviewing the manner in which fields can become massive with couplings to the conformal compensator \cite{Randall:1998uk}. For instance, consider a chiral superfield $S$ with a bilinear coupling in the K\"ahler potential. 
\be
\int d^4 x\> d^4 \theta \> \lambda \phi^\dagger \phi S^2.
\ee
Upon rescaling by $S \rightarrow \phi S$ we have
\be
\int d^4 x\> d^4 \theta \> \lambda \frac{\phi^\dagger}{\phi} S^2
\ee
Since $S^2$ is a chiral superfield, only terms involving the $F$-component of the conformal compensator appear in the Lagrangian. In this example,  assuming $\phi=1+m_{3/2}\theta^2$\cite{foot}, we are left with
\be
\int d^4 x \> \left( \int d^2 \theta \frac{\lambda}{2} m_{3/2} S^2 + h.c. \right) +\frac{ \lambda}{2} m_{3/2}^2 s^2.
\ee
Since we roughly identify the weak scale with $m_{3/2}/16 \pi^2$, $S$ naturally has a mass of order $10\ \tev$.

The low energy phenomena of the theory depend critically on the value of $\lambda$. The scalar mass matrix for $S$ is
\be
\frac{m_{3/2}^2 }{2}
\pmatrix{
\lambda^2 & \lambda \cr
\lambda & \lambda^2
}.
\ee
For $|\lambda| >1$ the mass matrix has positive determinant, and the situation is straightforward. However, with $|\lambda |<1$, this matrix has a negative eigenvalue, and $s$, the A-component of $S$, will  acquire a vev. The size of the vev will be determined by the presence of additional operators. If the superpotential contains a term $S^3$, the vev will naturally stabilize at the scale $m_{3/2}$.  One might think that if the vev is stabilized by higher dimension operators that it would become quite large, as in DAM. However, within this framework, the presence of fields with masses $O(m_{3/2})$ is quite natural, and hence upon integrating them out, one would be left with higher dimension operators suppressed by this mass scale. As a consequence, it is easy to find natural ways to stabilize the the vev of $s$ at $O(m_{3/2})$ .


This is our main point: {\em with a gravitino mass $m_{3/2}\sim O(10\ \tev)$, a natural scale for physics beyond the MSSM is $O(10\ \tev)$}. This physics could consist of new heavy particles, or a new symmetry breaking scale. Furthermore, in AMSB, threshold effects at 10 TeV or below do not decouple, and  the low energy superpartner mass spectrum is significantly changed. We will now examine two  attractive scenarios for  physics at this scale. We find that   AMSB need look very little like the minimal case. In some cases the spectrum is similar to  a variant of gauge mediation.

\subsection{N-viable Anomaly Mediation or Positive Deflection at $10\ \tev$}
We begin with the simplest example of additional  chiral superfields $\Xi,\overline \Xi$  in a nontrivial vector-like representation of the standard gauge group and the K\"ahler potential term
\be 
\int d^4 \theta \> \lambda  \phi^\dagger \phi \Xi\overline \Xi = \int d^2 \theta \lambda m_{3/2} \phi   \Xi\overline \Xi + h.c.
\ee
We assume $\lambda > 1$ to prevent color-charge breaking.

When we rescale the messenger fields $\phi \Xi \rightarrow \Xi$, the remaining superpotential term is
\be
\int d^2 \theta \frac{\lambda m_{3/2}}{\phi} \Xi \overline \Xi.
\ee
Notice that when we Taylor expand this in powers of $\theta$, the
$B\mu$ term arising from the conformal compensator now has the {\em
  opposite} sign when compared with the usual anomaly mediated
piece. These messengers will decouple in the usual fashion (up to
higher powers in $F/m^2$) when considering only the scalar masses
squared. However, their effects on the gaugino masses will remain. The
gaugino masses at the threshold are 
\be
M_i = -\frac{b_i + 2n}{4 \pi} \alpha_i m_{3/2}\ ,
\ee
where $n$ is the Dynkin index of the $\Xi$ fields and $b_i$
is the coefficient of the one loop beta function for $\alpha_i$.

Because the gaugino masses are no longer on the anomaly mediated trajectory, running from the threshold scale can change the masses of the sfermions considerably. Their soft masses at a scale $\mu < m_{\Xi}$ are given by
\bea
&m_{f}^2& =m_{3/2}^2 \times \hfil \\ \nonumber & \sum_i& -\frac{b_i c_{f,i} \alpha_i^2(m_\Xi)}{8 \pi^2} + \frac{c_{f,i} (b_i+2n)^2}{8 \pi^2 b_i}(\alpha^2_i(m_\Xi) - \alpha^2_i(\mu)),
\eea
where $i$ indexes the gauge group, $f$ indexes the sfermion
representation,and $c_{f,i}$ is the quadratic Casimir.  For $n\ge 5$, we obtain positive slepton masses squared. Notice that  the case $n=0$  reproduces the ordinary anomaly mediated result.

\subsection{$SU(3)$ electroweak at $10\ \tev$}
Possibly the most interesting way to solve  the negative slepton mass squared problem is to change the gauge structure of the MSSM matter fields at the $10\ \tev$ scale.  For instance, if we can place all the MSSM fields into asymptotically free gauge groups, their soft masses at $10\ \tev$ will be positive. In minimal anomaly mediation, enlarging the gauge group of the standard model above the scale $m_{3/2}$ is irrelevant due to decoupling, as only the gauge group structure at the weak scale  determines the soft scalar masses at the weak scale. However in EAM,  the  gauge contributions occur at the scale $m_{3/2}$ and  do not decouple. 

It is interesting  to extend the electroweak gauge group to include an additional $SU(3)_W$ factor, with all  the leptons transforming as triplets \cite{Konopinski:1953gq,Weinberg:1972nd}. 
 Dimopoulos and Kaplan \cite{Dimopoulos:2002mv}
showed that this naturally gives a successful prediction for the weak angle  when the extended gauge group is broken at a few TeV, and the $SU(2)\times U(1)$ couplings are fairly large.
Such an extension is attractive for EAM, since with the leptons transforming under an asymptotically free group the anomaly mediated contribution to their masses squared is positive.
 Here we  build an EAM  model with positive slepton masses squared and weak gauge group $SU(3)_W\times SU(2)\times U(1) $, which breaks to  $SU(2)_W\times U(1)_Y$ at a scale of order $m_{3/2} \sim$ 10 TeV.  The matter content is as follows. 
\begin{center}
{ {\bf  3 3 2 1 model Chiral Superfields}}
\end{center}
\underline{Quark, lepton and spectator sector (R parity odd)}
 \begin{equation}
\begin{tabular}{|r|c c c  c|}
\hline
  &$ SU(3)_c$& $SU(3)_W$ & $SU(2)$ &U(1) \\
\hline
$\ell_i$ & 1 & 3 & 1 & 0 \\
$q_i$& $3$ & 1 & 2 & 1/6\\
$\bar u_i$  &$\bar 3$ & 1 & 1 & -2/3 \\
$\bar d_i$ & $\bar 3$ & 1 & 1 & 1/3 \\ 
$\chi_2$ & 1 & $ \bar 3$ & 2 & -1/2\\ 
$\chi_1$ & 1 & $ \bar 3$ & 1 & 1 \\ 
\hline
\end{tabular}
\end{equation}
\underline{Higgs sector (R parity even)}
\begin{equation}
\begin{tabular}{|r|c c c  c|}
\hline
  &$ SU(3)_c$& $SU(3)_W$ & $SU(2)$ &U(1)  \\
\hline
$H$& 1& 1 & 2&1/2\\ 
$ \bar H$& 1 & 1 & 2 & -1/2\\ 
$\Sigma_2$ & 1 & $ \bar 3$ & 2 & -1/2\\
$\Sigma_1$&  1 & $ \bar 3$ & 1 & +1 \\
$\bar \Sigma_2$ & 1 & $  3$ & 2 & +1/2\\
$\bar\Sigma_1$ &  1 & $ 3$ & 1 & -1 \\ 
\hline
\end{tabular}
\end{equation}

Note that $i=1,2,3$ is a generation index.  The $\chi_{1,2}$ are spectators which obtain mass at the scale $m_{3/2}$, and for three generations, cancel the triangle gauge   anomalies  from  the quarks and leptons.

We include superpotential terms
\be
\phi^3  (g \Sigma_2^2\Sigma_1 +\bar{g} \bar \Sigma_2^2\bar \Sigma_1)
\ee
and  K\"ahler terms 
\be
\phi\phi^\dagger(\lambda_1 \Sigma_1\bar \Sigma_1+\lambda_2\Sigma_2 \bar \Sigma_2)
\ee
For simplicity, we set $\lambda_1=\lambda_2=\lambda$ and assume all couplings to be real. For $|\lambda|< 1$ the potential is unbounded from below along the direction where the  $3\times 3$ matrices $\Sigma=(\Sigma_1\ \Sigma_2)\propto (\bar \Sigma_1\ \bar \Sigma_2)=\bar \Sigma$ are rank one. However, for $\lambda<-1$ and $g,\ \bar g$ of the same sign, the global minimum can break $SU(3)_W\times SU(2)\times U(1)$ to 
$SU(2)_W\times U(1)_Y$ at a scale of order 10 TeV, with both $\Sigma$ and $\bar \Sigma$ proportional to the identity. If we relax the requirement that $\lambda_1=\lambda_2$, we will still break to $SU(2)\times U(1)$, in stages. Either  we have $SU(3)\times SU(2) \times U(1) \rightarrow SU(2)^2 \times U(1) \rightarrow SU(2)\times U(1)$ or$SU(3)\times SU(2) \times U(1) \rightarrow SU(2) \times U(1)^2 \rightarrow SU(2)\times U(1)$. Quantitatively, this will contribute only a threshold loop effect to the low energy value of $\sin^2 \theta_W$.

To compute the scalar masses, we  follow Giudice and Rattazzi \cite{Giudice:1998ni}. We treat each vev as a superfield with both $F$- and $A$- components, and extract the scalar masses from wave function renormalization. If the vev $X$ is driving the breaking $SU(3)\rightarrow SU(2)\times U(1)$, we parametrize $F_X/X $= $\gamma\> m_{3/2}$, we can write the mass of the sleptons as (considering only gauge interactions)
\bea
\nonumber m_{\tilde f}^2 = m_{3/2}^2 \times \Big(-\frac{\alpha_w^2 c_{f,w} b_w}{8 \pi^2} (1-\gamma)^2\\
-\frac{\alpha_Y^2 \gamma c_{f,Y}}{8 \pi^2} 
(b_Y \gamma + 2 b_Y' (1-\gamma) + 6 b_w (1-\gamma)) \\
\nonumber
- \frac{\alpha_2^2 \gamma c_{f,2}}{8 \pi^2} (b_2 \gamma + 2 b_2' (1- \gamma) 
+ 2 b_w (1-\gamma)) \Big)
\eea
where $\alpha_{Y,2}$ are the couplings of the low energy groups, and
$b_i'$ and $b_i$ are the beta functions of the $i$th gauge group above and 
below the $SU(3)$ breaking scale, respectively.
$\gamma=0$ corresponds to retaining the SUSY breaking masses of the theory above $m_{3/2}$ while $\gamma=1$ maintains the anomaly mediated trajectory, with negative slepton masses squared. For sufficiently small (but still order one) $\gamma$, the scalar masses squared will be positive. Radiative effects from the top-stop loops will drive electroweak symmetry breaking via the vev of $H$.
Masses for quarks, leptons and spectators arise from the superpotential couplings
\bea
\phi^3\Bigg(&& \!\!\!\!\!\!\frac{1}{m_{3/2}^2} h^l_{ij} \ell_i \ell_j \Sigma_2 \Sigma_1 \bar H
+ h^u_{ij} q_i\bar u_j H+h^d_{ij} q_i\bar d_j \bar H  \cr
&&+h_1 \chi_2^2\Sigma_1+
h_{2}\chi_1\chi_2\bar\Sigma_2+\xi_i\chi_2\ell_i H\Bigg)\  .
\eea
The nonrenormalizable operators giving rise to lepton masses can
arise, for instance, from integrating out vector-like leptons charged
under 
$SU(2)\times U(1)$ at the scale $m_{3/2}$. The fields responsible for
the lepton Yukawas will affect the gaugino masses but not the scalar
masses at leading order in supersymmetry breaking, as described in the
previous section. The light lepton fields will mix weakly with the
vectorlike leptons carrying 
$SU(2)\times U(1)$, but this need not lead to large corrections to the masses.

Note that the neutral $\chi_2$ fields can mix with the neutrino
components of the lepton fields via the $\xi_i$ couplings---this could
lead to lepton number and  flavor violation in the neutrino sector.   
For $\xi_i=0$, a linear combination of a global symmetry and
an $SU(3)_W$ generator remains unbroken in the ground state, and can
be identified as total lepton number. It would be interesting to study
whether the observed
 neutrino oscillations can be accounted for in this model.

\section{The Higgs $\mu$ parameter}

So far we have avoided dwelling on a potential embarrassment for EAM, the fact that in the MSSM the Higgs fields are in a real representation of the MSSM gauge group and so, according to our philosophy, should either remain massless or obtain masses of order 10 TeV. The most popular solution in the MSSM is to simply allow a bilinear Higgs term to appear in the K\"ahler potential \cite{Giudice:1988yz,Randall:1998uk,Nomura:2001ub}, which will give a Higgs $\mu$ parameter of order $m_{3/2}$. With $m_{3/2}\sim 10$ TeV this won't work. Even if one were to simply put in an  small parameter so that the $\mu$ parameter for the Higgs fields were in the phenomenologically desirable  range of several hundred GeV, the $B$ parameter would be  10 TeV and there would be no stable minimum of the potential.  
To get an acceptable spectrum, we therefore extend the MSSM at the TeV scale. The minimal such extension is to add a singlet $S$ to the MSSM,  with superpotential couplings
$\lambda SH_uH_d + k S^3$. $S$ must get an $\CO(100)$ GeV vev at the minimal of its potential. In minimal AMSB  the $S$ mass squared term can be computed to be positive. 
One way for getting $S$ a large negative mass squared, suggested in ref. \cite{Dine:1993yw} in the context of gauge mediation,
is to add new colored fields  $Q,\overline{Q}$ to the MSSM in a vector-like representation and with superpotential coupling
$h SQ\overline Q$. This mechanism will also work in anomaly mediation. Unless it is quite large, the  coupling constant $h$ is asymptotically free, and  will give $S$ a negative mass squared.

\section{Summary}

In anomaly mediated theories of supersymmetry breaking, 10 TeV is a natural mass scale for new physics beyond the MSSM. Such new physics makes a strong imprint on the superpartner spectrum, but can naturally preserve flavor universality. We have presented two simple models for new physics, and shown that the resulting superpartner mass spectrum is acceptable, and resembles none of the standard supersymmetry breaking scenarios. In particular, the gaugino masses are neither of the anomaly mediated nor unified type. In conclusion, anomaly mediation is a much richer framework than has been previously realized.

\par 
\vskip 0.1in \par
 
{\bf \noindent Acknowledgments}  
 
\vskip 0.05in 
 
\noindent 

This work was partially supported by DOE  contract DE-FGO3-96-ER40956.  We thank M. Luty, Y. Nomura, Z. Chacko and P. Fox for useful conversations, and A. Pierce for pointing out mistakes in an earlier version.
 
\bibliography{eam} 

\begin{thebibliography}{31}
\bibitem[{foo()}]{foot}
\bibinfo{journal}{Note that the F-component of $\phi$ is not always $m_{3/2}$,
  as in, e.g. \cite{Luty:2002ff}. What matters is not actually the gravitino
  mass but that $\phi|_F$ is of order 10 TeV.} .
\end{thebibliography}
 
\bibliographystyle{apsrev}

\end{document}